\documentclass[onecolumn]{aastex62}
\usepackage{amsmath}
\usepackage{amssymb}
\usepackage{mathtools}
\usepackage{multirow}

\providecommand{\deriv}{\ensuremath{\mathrm{d}}}
\providecommand{\given}{\ensuremath{\hspace{0.05em}\mid\hspace{0.05em}}}
\providecommand{\logten}{\ensuremath{\log_{\rm 10}}}

\providecommand{\gaia}{{\it Gaia}}
\providecommand{\gdr}[1]{GDR{#1}}
\providecommand{\gmag}{\ensuremath{G}}
\providecommand{\parallax}{\ensuremath{\varpi}}
\providecommand{\sigparallax}{\ensuremath{\sigma_{\varpi}}}
\providecommand{\fpu}{\ensuremath{\sigparallax/\parallax}}
\providecommand{\glon}{\ensuremath{l}}
\providecommand{\glat}{\ensuremath{b}}
\providecommand{\rlen}{\ensuremath{L}}
\providecommand{\rlengal}{\ensuremath{\rlen_{\rm gal}}}
\providecommand{\rlensph}{\ensuremath{\rlen_{\rm sph}}}
\providecommand{\dist}{\ensuremath{r}} 
\providecommand{\rest}{\ensuremath{\dist_{\rm est}}} 
\providecommand{\rlo}{\ensuremath{\dist_{\rm lo}}} 
\providecommand{\rhi}{\ensuremath{\dist_{\rm hi}}} 
\providecommand{\rmode}{\ensuremath{\dist_{\rm mode}}}

\received{26 April 2018}
\revised{30 May 2018}
\accepted{5 June 2018}
\published{20 July 2018}
\journalinfo{\href{http://iopscience.iop.org/article/10.3847/1538-3881/aacb21}{AJ 156, 58 (2018)}} 

\shorttitle{\gaia\ DR2 distances}
\shortauthors{Bailer-Jones et al.}

\begin{document}


\title{ESTIMATING DISTANCES FROM PARALLAXES IV:\\DISTANCES TO 1.33 BILLION STARS IN \gaia\ DATA RELEASE 2}

\correspondingauthor{Coryn A.L.\ Bailer-Jones}
\email{calj@mpia.de}

\author{C.A.L.\ Bailer-Jones}
\affil{Max Planck Institute for Astronomy, Heidelberg, Germany}

\author{J.\ Rybizki}
\affil{Max Planck Institute for Astronomy, Heidelberg, Germany}

\author{M.\ Fouesneau}
\affil{Max Planck Institute for Astronomy, Heidelberg, Germany}

\author{G.\ Mantelet}
\affil{Astronomisches Rechen-Institut, Zentrum f\"ur Astronomie der Universit\"at Heidelberg, Germany}

\author{R.\ Andrae}
\affil{Max Planck Institute for Astronomy, Heidelberg, Germany}



\begin{abstract}
For the vast majority of stars in the second \gaia\ data release, reliable distances cannot be obtained by inverting the parallax.
A correct inference procedure must instead be used to account for the nonlinearity of the transformation and the asymmetry of the resulting probability distribution. Here we infer distances to essentially all 1.33 billion stars with parallaxes published in the second \gaia\ data release. This is done using a weak distance prior that varies smoothly as a function of Galactic longitude and latitude according to a Galaxy model. The irreducible uncertainty in the distance estimate is characterized by the lower and upper bounds of an asymmetric confidence interval. Although more precise distances can be estimated for a subset of the stars using additional data (such as photometry), our goal is to provide purely geometric distance estimates, independent of assumptions about the physical properties of, or interstellar extinction towards, individual stars. We analyse the characteristics of the catalogue and validate it using clusters.
The catalogue can be queried on the Gaia archive using ADQL
at \url{http://gea.esac.esa.int/archive/} 
and downloaded from \url{http://www.mpia.de/~calj/gdr2_distances.html}.
\end{abstract}



\section{Introduction}\label{sec:introduction}

Essentially all published quantities are inferences as opposed to direct measurements. Even apparently simple quantities such as celestial coordinates are achieved only after extensive data processing, e.g.\ reducing CCD images, fitting a point spread function, applying a focal plane geometric calibration, transforming to a global coordinate system. 
The role and impact of the assumptions involved in this process should not be overlooked.

The parallaxes published in the second \gaia\ data release (hereafter \gdr{2}; \citealt{DR2-DPACP-36}) are obtained after a complex iterative procedure, involving various assumptions \citep{2012A&A...538A..78L}. Going from a parallax to a distance is also a non-trivial issue, as has been described in several previous works (see \citealt{DR2-DPACP-38} for a recent discussion). The main issues are the nonlinearity of the transformation and the positivity constraint of distance (but not the parallax).
Many parallaxes in \gdr{2} also have very low signal-to-noise ratios (this will be quantified later), meaning that a characterization of the inferred distance uncertainties is at least as important as a single point estimate of the distance. 

The only consistent and physically meaningful way to infer distances and their uncertainties is through a probabilistic analysis, as described for example in \cite{2015PASP..127..994B}. This involves specifying a likelihood and a prior. 
The likelihood is defined by the parallax inference process, and on account of the assumptions involved, the linearity of the model, and the central limit theorem, this is theoretically a Gaussian distribution (L.\ Lindegren, private communication), 
and it has also been confirmed empirically that this is a very good approximation for \gdr{2} \citep{DR2-DPACP-51}.  
The prior should incorporate any relevant aspects of the expected distance distribution that are not contained within the likelihood. The obvious one is the positivity of distance. Others might be the properties of the survey and our current knowledge of the Galaxy.

What kind of prior one adopts depends in part on the desired trade-off between model-dependence and precision: a complex prior based on a detailed Galaxy model will follow this more closely in the limit of poor data, which may or may not be what one wants.  As such one should not single out priors for introducing ``biases'', as their use is conceptually no different from what is done in any inferential process, which always involves a combination of both ``measurements'' and assumptions. Indeed, there is no such thing as uninformative priors or model-free inference. The issue is not whether a bias is present, but what impact our choices have.  The advantage of the probabilistic approach is that it combines the measurement (likelihood) and assumptions (priors) in a consistent way that makes the prior irrelevant when the data are good, but ensures a graceful transition to dominance by the prior as the data quality degrades. 

\cite{2015PASP..127..994B} and \cite{2016ApJ...832..137A} (papers I and II in this series) investigate the consequences of using different types of priors, including one based on a Galaxy model.  Using the first \gaia\ data release of two million parallaxes, \cite{2016ApJ...833..119A} (paper III) inferred distances using priors at two extremes: (1) a simple isotropic prior; (2) a prior 
given by the distribution of stars along each line-of-sight as determined from a Galaxy model, which also accounted for interstellar extinction and the \gaia\ selection function.

In this paper we infer distances to the 1.33 billion sources in the \gdr{2} catalogue that have parallaxes. We adopt
the exponentially decreasing space density (EDSD) prior in distance \dist
\begin{equation}
P(\dist\given\rlen)  \ = \  \begin{dcases}
  \ \frac{1}{2\rlen^3}\,\dist^2e^{-\dist/\rlen}  & \:{\rm if}~~ \dist >0 \\
  \ 0                          & \:{\rm otherwise}
\end{dcases}
\label{eqn:prior1}
\end{equation}
where $\rlen>0$ is a length scale, as described in \cite{2015PASP..127..994B} and \cite{2016ApJ...832..137A}, and adopted by \cite{2016ApJ...833..119A} for \gdr{1}. This prior has a single mode at $2\rlen$. In the current paper \rlen\ varies as a function of longitude and latitude $(\glon,\glat)$
according to a model, to reflect the expected variation in the distribution of stellar distances in the \gaia-observed Galaxy. This is, in some sense, a compromise between the simplicity of the isotropic prior (i.e.\ fixed \rlen) and the complexity of the line-of-sight-dependent distribution shapes obtained from a Galaxy model.

In this work we only use the \gaia\ astrometry and our length scale model to infer distances. Given the limited fractional precision of the parallaxes for more distant stars, this limits the distance precision we can achieve.
One could, of course, use other information to improve the distance estimates further. An obvious addition is to use a model for the absolute magnitude of the star together with a measurement of its apparent magnitude
and a measurement of (or model for) the line-of-sight extinction. An absolute magnitude model is normally based on
the Hertzsprung--Russell diagram and the observed colours or spectroscopy of the star \cite[e.g.][]{2016ApJ...832..137A}. Such an approach was taken by \cite{2017arXiv170704554M} to estimate distances using the overlap between RAVE
 spectroscopy and the two million parallaxes in \gdr{1}. Other recent studies combining parallaxes with other information to improve distance measures include 
\cite{2014MNRAS.443..698S}, \cite{2017AJ....154..222L}, \cite{2017ApJ...838..162O}, \cite{2017arXiv170605055A}, \cite{2018MNRAS.476.2556Q}, \cite{2018arXiv180110042C}, \cite{coronado2018}, and Fouesneau et al.\ (in preparation).

Our goal here is different. We choose to infer distances without
invoking any assumptions about the properties of, or the extinction towards, individual stars.
(Stellar properties and extinction only enter in a collective sense to construct the prior.)
This enables us to produce a self-consistent
catalogue for all of \gdr{2}. The price we pay is precision. Our catalogue provides a purely geometric measure of distance and its uncertainty that overcomes the dangers of inverse parallax and unphysical priors. The obvious use case for this catalogue is to provide the distance to one or more stars, or rather the probable range of distances to the stars, since the uncertainties are real and often significant. This can tell us, for example, whether a set of stars have consistent distances.  The catalogue can be used to select a sample of stars on which other inferences are then performed, or for which more precise distances are estimated using additional information. The catalogue distances also serve as a baseline against which to compare other distance estimates. We may likewise use the catalogue to determine the three-dimensional space distribution of a set of stars. Note that although the distances are determined independently from one another, the prior is correlated on small spatial scales. Thus if we know or suspect that the stars are members of a stellar cluster, a combination of our distances may not provide a reliable distance to the cluster.  It is far better to set up a model for the cluster in which its distance is a free parameter, and to solve for this using the original parallaxes (accommodating also their spatial correlations). Generally speaking it is suboptimal to use our distances if they are just an intermediate step in a calculation, e.g.\ if one really wants to estimate absolute magnitudes or the transverse velocity of a cluster. More accurate results (and a better propagation of uncertainties) will be obtained by inferring directly the quantities of interest.

In the next section we describe the prior model and the computation of distances and asymmetric uncertainties.
In section~\ref{sec:results} we analyse the results. The content and characteristics of the catalogue are presented in section~\ref{sec:catalogue}. We summarize in section~\ref{sec:summary} with some more notes on how (not) to use the catalogue.

\section{Method}\label{sec:method}

\subsection{Posterior probability density function}

Given the Gaussian likelihood in the parallax \parallax\ with standard deviation \sigparallax, plus the EDSD prior from equation~\ref{eqn:prior1}, the unnormalized posterior over the distance to a source is 
\begin{equation}
P^*(\dist\given\parallax, \sigparallax, \rlensph(\glon,\glat))  \ = \  \begin{dcases}
  \ \dist^2 \exp{ \left[ -\frac{\dist}{\rlensph(\glon,\glat)} -\frac{1}{2\sigparallax^2}\left(\parallax - \parallax_{\rm zp} -\frac{1}{\dist}\right)^2 \right] }  & \:{\rm if}~~ \dist > 0 \\
  \ 0                          & \:{\rm otherwise} \ .
\end{dcases}
\label{eqn:post1}
\end{equation}
The quantities \parallax, \sigparallax, \glon, \glat\ are all taken from the 
{\tt gaia\_source} table in \gdr{2} from the columns {\tt parallax}, {\tt parallax\_error}, {\tt l}, {\tt b} respectively.
$\parallax_{\rm zp}$ is the global parallax zeropoint, determined from \gaia's observations of quasars to be $-0.029$\,mas \citep{DR2-DPACP-51}. (The zeropoint actually varies as a function of sky position, magnitude, and colour, but this is the best global estimate.) The length scale $\rlensph(\glon,\glat)$ has a Galactic longitude and latitude dependence as described in section~\ref{sec:lengthscale}. 

The properties of this posterior have been explored in some detail in \cite{2015PASP..127..994B}. For physical values of its three parameters -- finite \parallax, positive \sigparallax, positive \rlensph\ -- 
it is always a proper (i.e.\ normalizable) density function. It can be either unimodal or bimodal, although for the vast majority of cases in \gdr{2} it is unimodal.

\subsection{Distance estimators}\label{sec:estimators}

While the posterior in equation \ref{eqn:post1} is the complete description of the distance to the source (fully specified by
 \parallax, \sigparallax\ in \gdr{2} and \rlensph\ in our catalogue),  we sometimes want a point estimate along with some measure of the uncertainty.
As known from previous work and seen from the examples in section \ref{sec:results}, this posterior is asymmetric, with a longer tail towards large distances. Being a proper (normalizable) posterior, all of the usual estimators -- mean, median, mode, variance, quantiles, full-width at half-maximum -- are defined. 
As the point estimator, \rest, we prefer here the mode, \rmode. This is found analytically by solving a cubic equation \citep{2015PASP..127..994B}.

As a measure of uncertainty we use the highest density interval (HDI) with probability $p$. The HDI is the span of the distance that encloses the region of highest posterior probability density, the integral of which is $p$.  The span is defined by the lower and upper bounds, \rlo\ and \rhi\ respectively.
Here we set $p=0.6827$, which is equal to the probability contained within $\pm 1 \sigma$ of the mode for a Gaussian distribution. The distance posterior is asymmetric (sometimes significantly so) so $\rhi - \rest$ and $\rest - \rlo$ are unequal. Conceptually, the HDI can be found by lowering a horizontal line over the distribution until the area contained under the curve between its intercepts with the curve (i.e.\ \rlo\ and \rhi) is equal to $p$ (see Figure 5.10 of \citealt{2017pbi..book.....B}).  
The HDI has the advantage over other confidence intervals in that it is guaranteed to contain the mode. 
Some other common measures are less flexible. The Fisher information approach, for example, assumes the posterior is locally Gaussian, the standard deviation of which is taken as the uncertainty (which is symmetric and therefore entirely inappropriate for positively-constrained quantities like distances).

This HDI is unique if the posterior is unimodal.
(A more general way of finding the HDI allows it to split into multiple intervals in the case of multimodality; see for example \citealt{hyndman1996}).
For a small part of the parameter space defined by $(\parallax, \sigparallax, \rlen)$, the posterior is bimodal (see \citealt{2015PASP..127..994B}). For these cases we retain both the mode estimator and the HDI if possible (i.e.\ if the span does not include the minimum, thereby retaining the uniqueness; see below). 
Otherwise we resort to using the median of the distribution as the point estimator, and report 
the 16th and 84th percentiles (i.e.\ $(1 \pm p)/2$) as \rlo\ and \rhi\ respectively; together these two numbers form the equal-tailed interval (ETI),  which has as much probability below the span as above, with $p$ in between. 

There is no analytic solution for the HDI for this posterior. We compute it with an iterative procedure involving a Taylor expansion.
The basic idea is to take a small step in both directions away from the mode, compute the area under the curve covered by this step, and iterate this until the total area hits the desired limit.

We first normalize the posterior using Gaussian quadrature; denote this with $P(\dist)$.
With \dist\ set equal to the mode (where the first derivative is zero), and adopting a fixed negative $\Delta P$, an initial step of size
\begin{equation}
\Delta\dist_0 = \ \sqrt{2\Delta P \left(\frac{\deriv^2P}{\deriv\dist^2}\right)_{\rmode}^{-1}}
\end{equation}
is computed. 
Candidate lower and upper bounds are then defined as 
\begin{equation}
\dist^{\pm}_1 \ = \ \rmode \pm \Delta\dist_0 \ .
\end{equation}
The area under the curve between these bounds is
computed using the trapezium rule. Assuming this area is less than $p$ ($\Delta P$ is set small enough to ensure this is always the case), further steps are computed in both the negative and positive directions using the first order term in the Taylor expansion, i.e.\
\begin{equation}
\Delta\dist^{\pm}_i \ = \ \Delta P \left(\frac{\deriv P}{\deriv\dist}\right)_{\dist^{\pm}_i}^{-1} 
\label{eqn:deltadist}
\end{equation}
starting from $i=1$, 
to give  new candidate lower and upper bounds as
\begin{equation}
\dist^{\pm}_{i+1} \ = \ \dist^{\pm}_i + \Delta\dist^{\pm} \ .
\end{equation}
The additional area covered by these steps is estimated using the trapezium rule and added to the running total area. This process is iterated from equation \ref{eqn:deltadist}, with the result that both bounds move monotonically away from the mode. The procedure is stopped when the running total area computed exceeds $p$ (or bimodality is detected; see below).
For the distance catalogue produced here, $\Delta P$ is fixed to $-0.01P(\rmode)$.
For 98\% of the sources in \gdr{2}, between 39 and 73 iterations (the 1st and 99th percentiles) are required.

Note that the area computed by this method actually exceeds $p$, although generally by only a very small amount: for 99.9\% of the sources the area computed is still less than 0.6973.
This is an acceptable degree of approximation. (Tests shows that the area calculation by the trapezium rule is generally very accurate: less than 0.06\% of cases deviate from an accurately-calculated area by more than 1 part in 1000.)
In a few pathological cases the posterior has a long, nearly flat tail extending to large distances. The numerical estimation of the HDI is then poor. In a very small fraction of cases this can even lead to the computed area exceeding 0.9. If this happens the mode and HDI are rejected as catalogue entries, and replaced with the median and equal-tailed quantiles, as explained below. 
(These solutions have {\tt result\_flag=2} and {\tt modality\_flag=1}. A full description of the flags is given in the next section.)

Because the steps in the positive and negative directions are made independently, the values of the posterior at \rlo\ and \rhi\ are not identical: an imaginary line being lowered over the posterior does not remain exactly horizontal.
Because $P=0$ only for $\dist=0$ and $\dist\ = +\infty$ (equation~\ref{eqn:post1}), and because \rhi\ can never reach infinity, then in principle \rlo\ should never reach exactly zero. With this numerical method it can, although it only happens in 183\,765 (0.014\%) cases.
The algorithm prevents steps to negative distances, so $\rlo \geq 0$.
(Note that this method of finding the HDI would not work without modification for distributions which drop to zero at finite values of $r$ greater than zero.)

If the posterior is bimodal (which occurs in 0.09\% of cases), then the search for the HDI is done starting from the highest mode. Provided the search does not encounter the minimum between the two modes, the HDI remains well defined, and this, together with the highest mode, are the posterior summary provided in the catalogue.
The {\tt result\_flag} has value {\tt 1} whenever the point estimate, \rest, is the mode, and the
lower and upper bounds of the confidence interval, \rlo\ and \rhi, define the HDI.

The presence of bimodality (which is identified independently of the search, from the roots of $\deriv P/\deriv \dist=0$) is always indicated by the {\tt modality\_flag}, which is {\tt 1} for unimodal and {\tt 2} for bimodal.
The minimum is encountered during the HDI search
in less than 0.1\% of all cases. If this happens we stop the search and instead compute the median (the 50th percentile) and the ETI, which 
we report 
in the catalogue.
Such solutions are indicated with {\tt result\_flag=2}.
The quantiles are found numerically by drawing $2\times10^4$ samples from the posterior with a Markov Chain Monte Carlo (MCMC) method.
Depending on the shape of the posterior, these estimates of the quantiles may not be very accurate, so should be treated with caution.

The HDI is a couple of orders of magnitude faster to compute than the ETI, because the former involves that many fewer calculations of the posterior density.

In a very small number of cases (about two in every million), the algorithm fails to give any summary of the posterior.
In these cases {\tt result\_flag=0} and the three distances \rest, \rlo, \rhi\ are set to {\tt NaN}.
This occurs when the parallax is negative yet highly statistically significant,  $\sigparallax/\parallax \ll -1$,
which results in the unnormalized posterior density being numerically equal to zero everywhere on account of the finite machine precision. This precludes both the computation of the normalization constant (for the HDI) 
and an MCMC sampling (for the quantiles). Although we can still compute the mode (it is found algebraically),
we choose not to report it. One could work around this problem by including scaling factors in the computation of the density, but the cases are very rare, plus the distance posterior is virtually identical to the prior, which is entirely specified by the length scale (always provided).

We publish only a single point estimate together with the bounds of a single confidence interval in order to keep the catalogue manageable in size. Should users want other estimates, e.g.\ different size confidence intervals, it is quick and easy to recompute the posterior using \parallax\ and \sigparallax\ in \gdr{2} and \rlensph\ in our catalogue. Processing all 1.33 billion sources took 75 CPU-core days, which on our multicore machine was just a few hours of wall-clock time. 
The code for the processing is available at {\url https://github.com/ehalley/Gaia-DR2-distances}.

\subsection{Length scale model}\label{sec:lengthscale}

We use a chemo-dynamical model to simulate the Galaxy (including extinction) as seen by \gaia, which here means all stars with apparent magnitude $G \leq 20.7$\,mag. The resulting mock catalogue and the Galaxy model on which it is based are described in \cite{2018arXiv180401427R}.  From this catalogue we extract the distances to all stars in each of the 49\,152 equal-sized (0.84\,square degrees) healpix cells (level 6) over the entire sky.  We then fit the EDSD prior, equation~\ref{eqn:prior1}, to the stars in each cell. The maximum likelihood estimate for \rlen\ is just $\overline{\dist}/3$, where $\overline{\dist}$ is the mean of the distances in that cell. To avoid a few stars at very large distances dominating this estimate we replace the mean with the median.

\begin{figure}
\begin{center}
\includegraphics[width=\textwidth, angle=0]{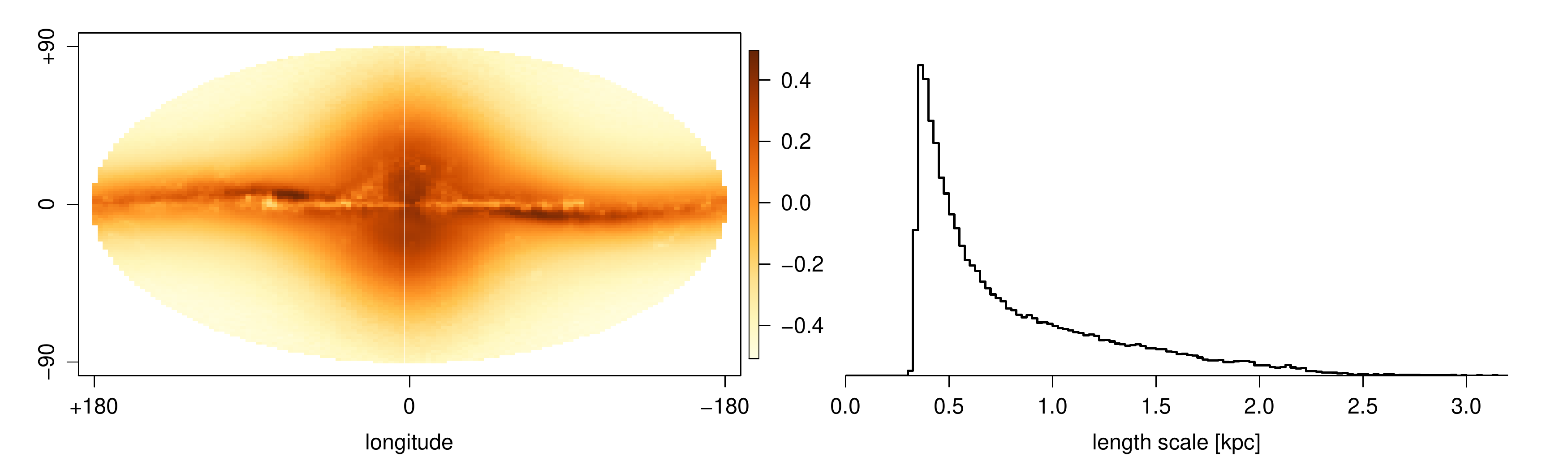}
\caption{Distribution of the length scales from the Galaxy model, $\rlengal(\glon, \glat)$. Left: The log (base 10) length scale in kpc shown in Galactic coordinates in a Mollweide (equal-area) projection. Right: Histogram (linear units on both axes) of the length scales (over equal-area cells).
The full range of \rlengal\ in the model (and covered by the colour scale) is 0.310 to 3.143\,kpc ($-$0.508 to 0.497 on the log scale).
\label{fig:rlen_galmod_galproj}
}
\end{center}
\end{figure}

The resulting map, $\rlengal(\glon,\glat)$, is shown in Figure~\ref{fig:rlen_galmod_galproj}.
The bulge, disk, and warp of the disk are quite prominent. Perhaps counter-intuitively, the median distance is generally smaller at high Galactic latitudes than at low latitudes. This is because of the very high density of stars far away in the Galactic disk which \gaia\ can see, despite the extinction. Although the maximum distance \gaia\ can observe to is larger at high latitudes, the density of star drops rapidly with height above the disk, so the median is determined primarily by nearer stars. The same map computed using $\overline{\dist}/3$ exhibits larger distances than this map at high latitudes.

\begin{figure}
\begin{center}
\includegraphics[width=\textwidth, angle=0]{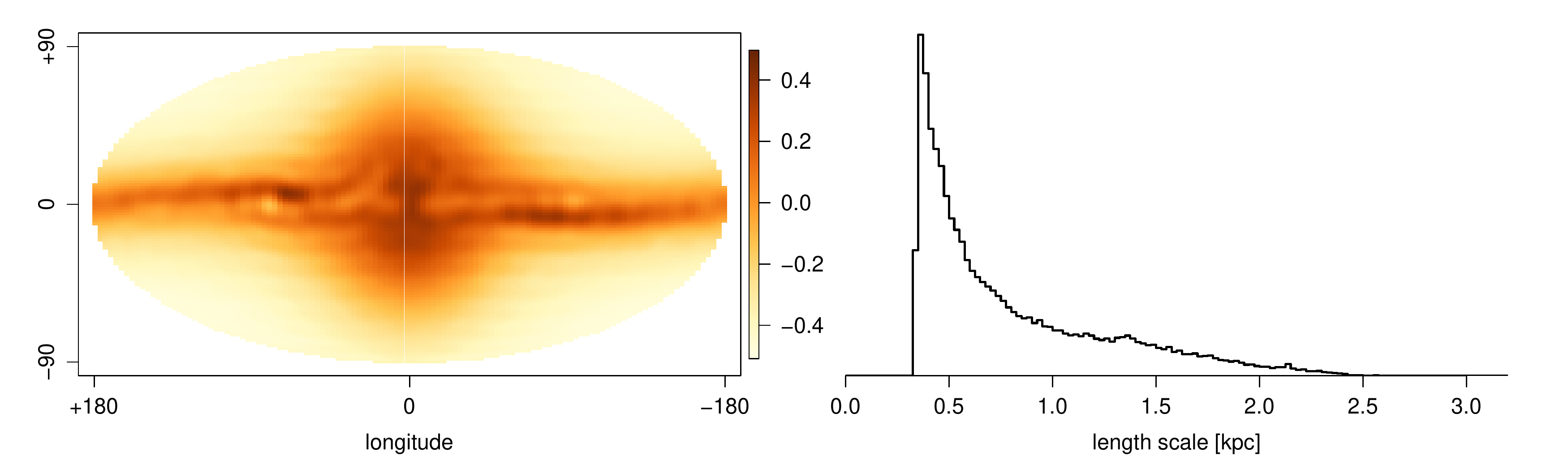}
\caption{Distribution of the spherical harmonic model fit to the length scales, $\rlensph(\glon, \glat)$. Left: The log (base 10) length scale in kpc shown in Galactic coordinates in a Mollweide projection. Apparent discontinuities in the map are due to the finite-sized cells used for plotting; they are not a property of the fit. 
Right: Histogram (linear units on both axes) of the length scales (over equal-area cells). The plotting ranges of the distance in both panels (colour scale and histogram) are the same as in Figure~\ref{fig:rlen_galmod_galproj}. The range of \rlensph\ values in the catalogue is given in Table~\ref{tab:catminmax}.
\label{fig:rlen_sphharm_galproj}
}
\end{center}
\end{figure}

\begin{figure}
\begin{center}
\includegraphics[width=0.5\textwidth, angle=0]{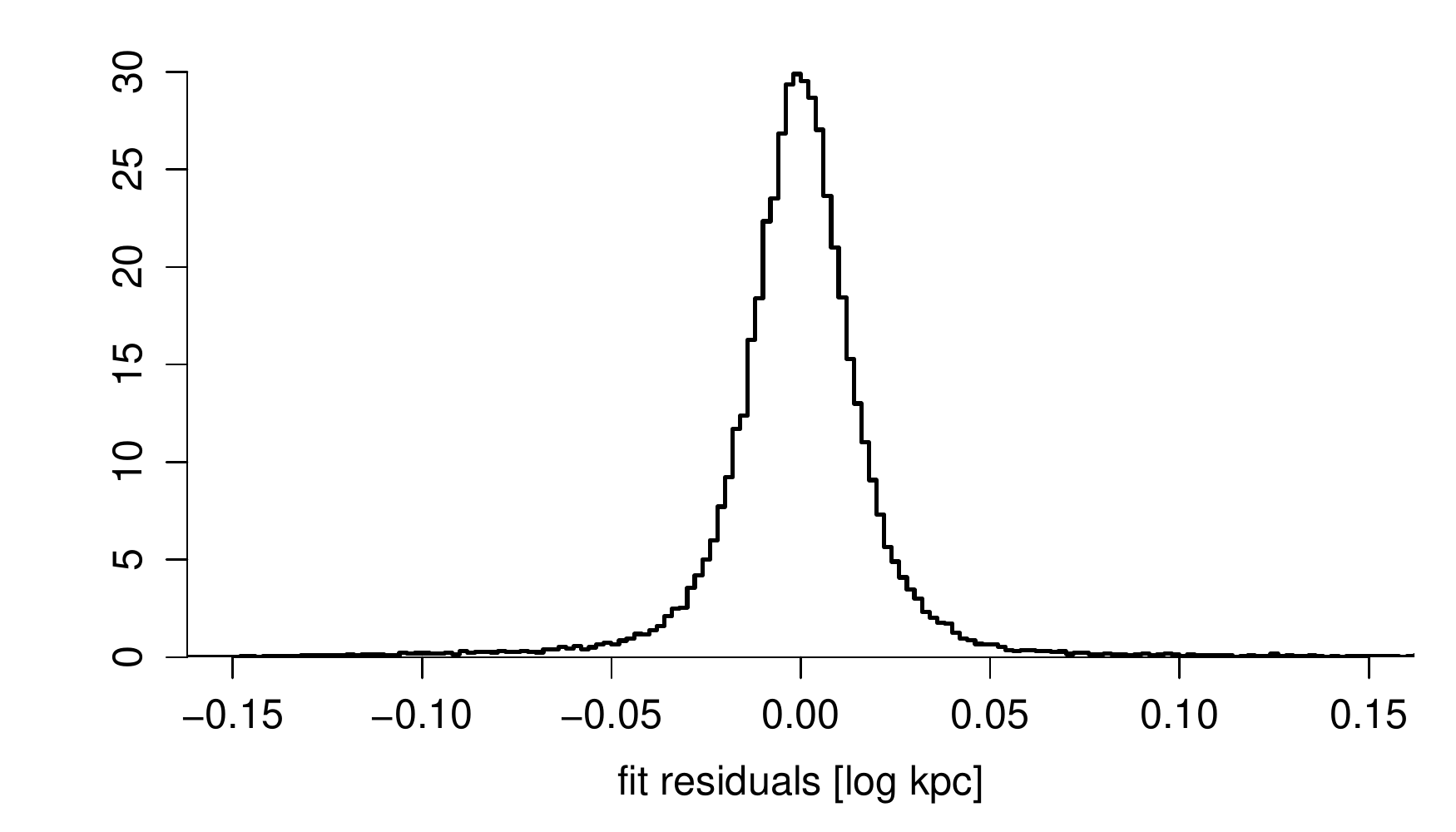}
\caption{
Histogram of the residuals of the fit of the spherical harmonic model, in the sense of $\logten\rlensph - \logten\rlengal$.
\label{fig:rlen_residuals_galproj}
}
\end{center}
\end{figure}

We do not use this map directly as the prior model, because it shows discontinuities across the healpix boundaries. This would lead to spatial discontinuities in the inferred distances. We therefore fit a spherical harmonic model of order $n_{\rm max}$
\begin{equation}
f(\glon,\glat)_{n_{\rm max}} \ = \  a + \sum_{n=1}^{n_{\rm max}}\sum_{m=0}^{n} s_{n,m}P_n^m(\sin \glat)\sin(m\glon) + c_{n,m}P_n^m(\sin \glat)\cos(m\glon)
\end{equation}
where $P_n^m()$ is the associated Legendre polynomial of order $(n,m)$, and $\{s_{n,m}, c_{n,m}\}$ together with $a$ are the free parameters to be determined.
Note as $P_n^m(\sin b)\sin(ml) = 0$ when $m=0$, the constants $s_{n,0}$ are unconstrained, so we set them to zero. The total number of free parameters is
\begin{equation}
1+ \sum_{n=1}^{n_{\rm max}}(2n+1) = (n_{\rm max} + 1)^2.
\end{equation}
We fit this model to the map of $\logten\rlen$ using linear least squares.  After some experimentation, also with the healpix level of the input map, we settled on $n_{\rm max} = 30$ as high enough to capture the essential spatial variations, but low enough to smooth out some of the rather sharp features in the model. 
The fitted map, which we denote $\rlensph(\glon,\glat)$, is shown in Figure~\ref{fig:rlen_sphharm_galproj}.
A histogram of the residuals of the fit is shown in Figure~\ref{fig:rlen_residuals_galproj}. 

\section{Analysis of distance results}\label{sec:results}

We now analyse the results of the distance inference, using a random subset of a million sources.

\subsection{Parallax statistics}

\begin{figure}
\begin{center}
\includegraphics[width=\textwidth, angle=0]{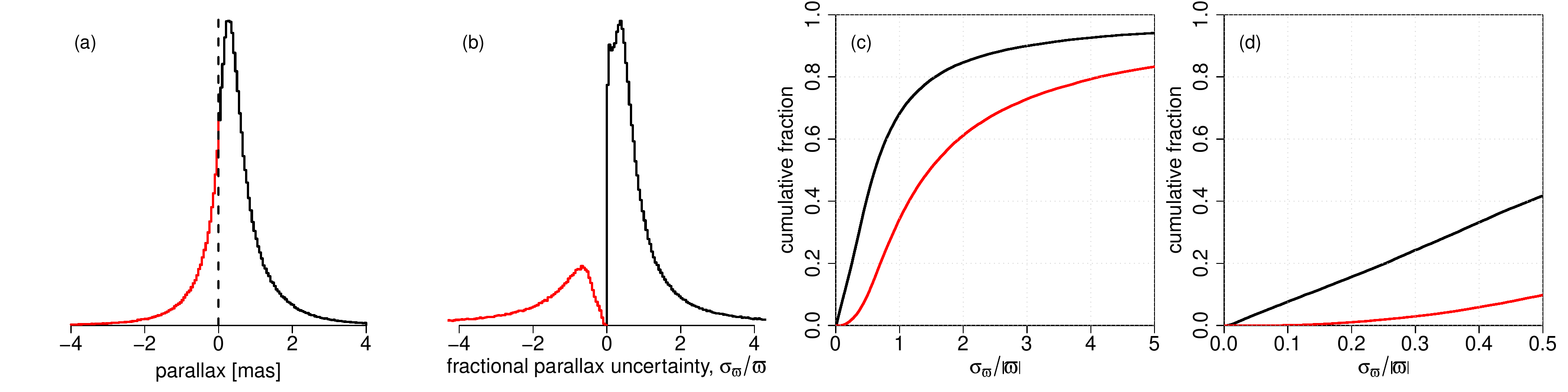}
\caption{The parallaxes and parallax precisions in \gdr{2}. (a) The distribution of parallaxes in \gdr{2}. (b) The distribution of fractional parallax uncertainties, \fpu.
Linear scales are used on both axes of the histograms.
This is shown as a cumulative distribution in panel (c) split for the positive parallaxes (black/upper line) and the negative parallaxes (red/lower line) (i.e.\ the two lines asymptote to values which sum to 1.0). Panel (d) is a zoom of (c).
\label{fig:parallax_distributions}
}
\end{center}
\end{figure}

The distribution of the parallaxes is shown in Figure~\ref{fig:parallax_distributions} (compare with the predictions in Figure~2 of \citealt{2016ApJ...832..137A}).
24.9\% of the sources have negative parallaxes. The distribution of the fractional parallax uncertainty, defined as \fpu, is shown in panel (b). Immediately striking is how many sources have very low signal-to-noise (SNR) parallaxes: of those sources with positive parallaxes, only 15.7\% have $\fpu<0.2$, i.e.\ have a SNR greater than 5 (7.7\% greater than 10, 0.98\% greater than 50). 
These negative and/or low SNR parallaxes are due to the numerical dominance of relatively distant and faint sources.
Cumulative distributions for \fpu\ are shown in panels (c) and (d) for positive (black curve) and negative (red curve) parallaxes.

\subsection{Example distance posteriors}

\begin{figure}
\begin{center}
\includegraphics[width=\textwidth, angle=0]{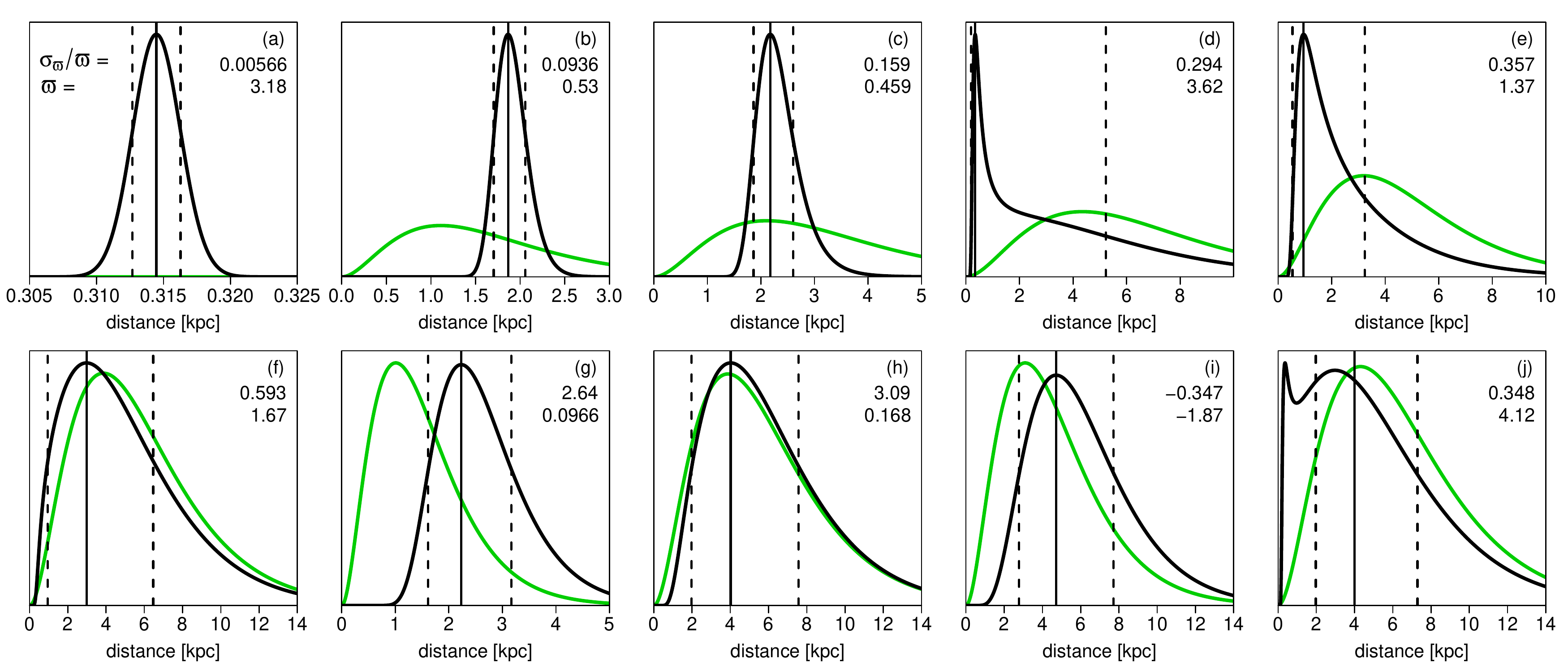}
\caption{Ten examples of the inferred posteriors (black curves) and the corresponding priors (green curves).
All distributions are normalized.
The two numbers in the top-right corner of each panel are the
fractional parallax uncertainty \fpu\ (upper), and the parallax in mas
(lower). Panels (a) to (i) are unimodal posteriors for which the
vertical solid line is the mode and the vertical dashed lines denote
the HDI containing 0.68 of the posterior probability. In panel (a) the
prior is so broad on this scale that it appears at almost zero
density. Panels (a) to (h) are ordered by increasing \fpu. Panel (i)
is for a negative parallax. Panel (j) shows a bimodal solution for
which the HDI cannot be computed: the solid line shows the median, and
the dashed lines show the 
ETI containing approximately 0.68 of the posterior probability.
\label{fig:posterior_prior_examples}
}
\end{center}
\end{figure}

Figure~\ref{fig:posterior_prior_examples} shows examples of several posteriors and their corresponding priors. (More examples can be found in Figure 12 of \citealt{2015PASP..127..994B}.) Recall that the mode of the posterior is at $2\rlensph$. We do not plot the likelihood because it is an improper distribution in $\dist$, 
so can only be plotted with an arbitrary scaling.
The exact shape of the posterior depends in a somewhat complex manner on the value of all three of its variables: \parallax, \fpu, and \rlensph.
These examples have been chosen to show a range of behaviours; they do not represent the frequency of such posteriors among the results. Panels (a) to (c) are for small positive values of \fpu: these posteriors are dominated by the likelihood (as opposed to the prior) and are therefore approximately Gaussian with a mode close to the inverse parallax. 
Panels (d) and (e) are cases where the likelihood and prior have similar size influence on the posterior. As \fpu\ increases, the positive distance tail extends and the posterior becomes increasingly asymmetric. Panel (d) is a relatively rare example of a posterior with a high, narrow peak accompanied by a long tail. 

Broadly speaking, the larger \fpu, the more the prior dominates the posterior. This is seen in panels (f) and (h). 
However, it is not simply the case that the larger the value of \fpu, the closer the posterior is to the prior.  We see this in panel (g), where \fpu\,=\,2.64 (larger than in panel f), yet the posterior mode is at more than twice the value of the prior mode. The posterior is always the (normalized) product of prior and likelihood, so if these peak at very different values, as is the case in panel (g) ($2\rlensph$\,=\,1.0\,kpc vs.\ $1/\parallax$\,=\,10.4\,kpc), the likelihood can still play a significant role, even when the data are poor (i.e.\ \fpu\ is large). 
In other words, prior dominance does not increase monotonically with \fpu. The actual values of the parallax and the prior length scale are also important.

Panel (i) is an example of a negative parallax. 
Panel (j) shows the relatively rare case of a bimodal posterior. In this particular case the HDI cannot be uniquely defined, so our algorithm returns the median as the point estimate, and the equal-tailed quantiles to define the asymmetric confidence interval.

\subsection{Distance statistics}

\begin{figure}
\begin{center}
\includegraphics[width=\textwidth, angle=0]{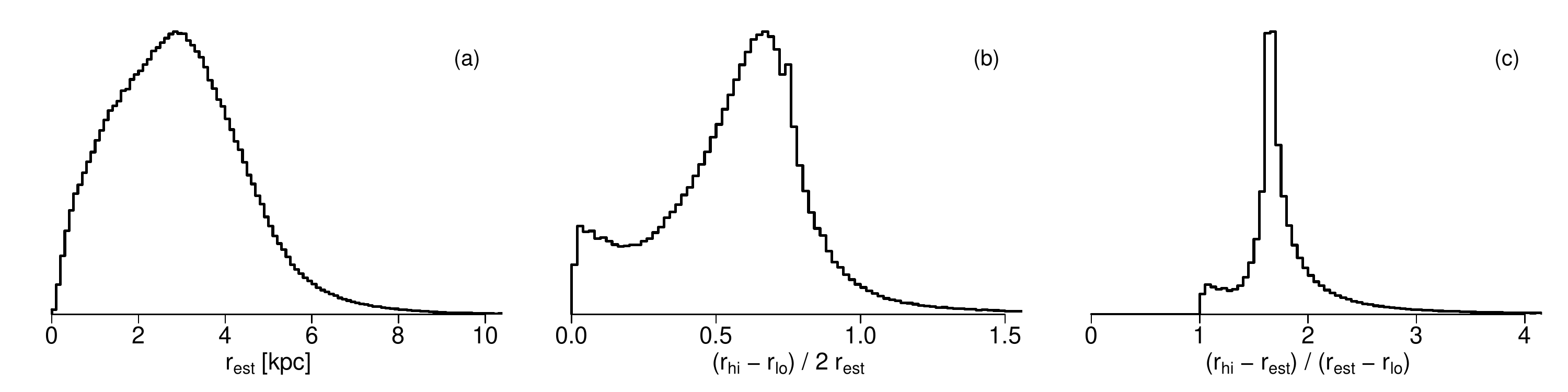}
\includegraphics[width=\textwidth, angle=0]{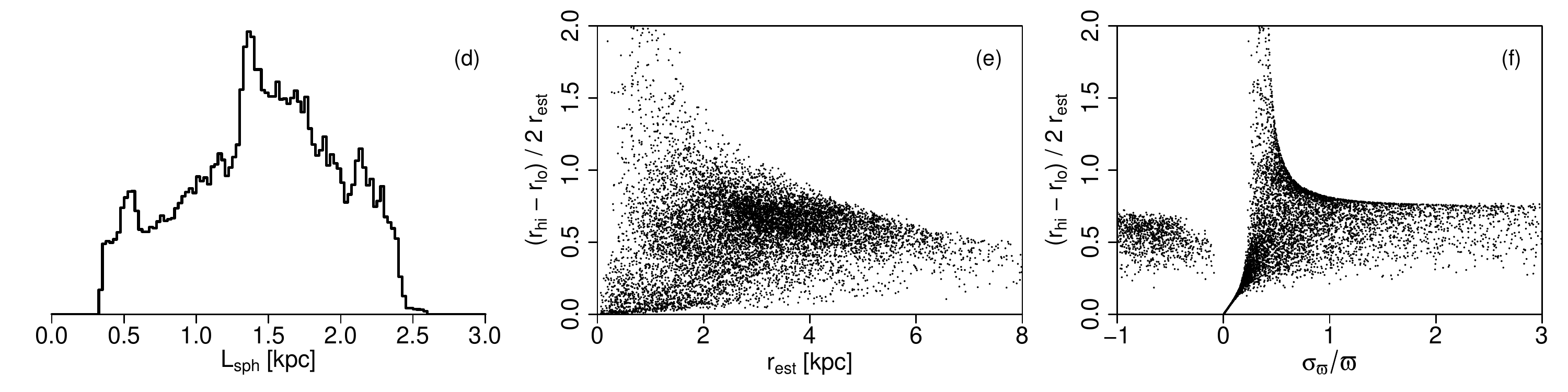}
\includegraphics[width=\textwidth, angle=0]{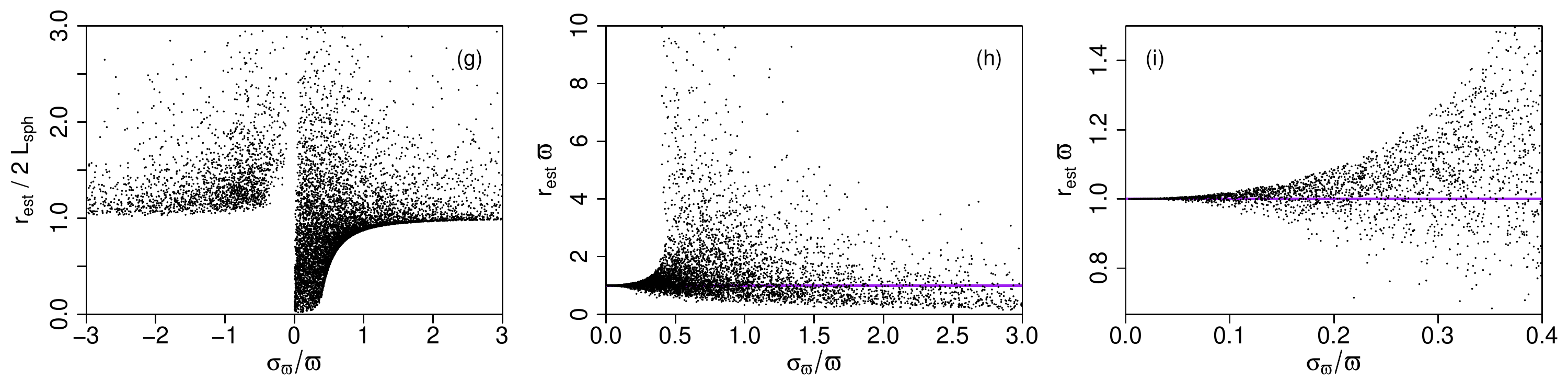}
\caption{Summaries of the distance inference for a randomly selected set of one million sources with 
{\tt result\_flag=1} and {\tt modality\_flag=1}.
Linear scales are used on both axes of the histograms.
To avoid crowding, panels (e)--(i) plot just 10\,000 of the sources. 
Note that panel (d) shows the distribution of \rlensph\ over stars, whereas the histograms of \rlen\ in Figures~\ref{fig:rlen_galmod_galproj}b
and~\ref{fig:rlen_sphharm_galproj}b are the density over equal-area cells on the sky.
In panels (h) and (i) -- the latter is a zoom of the former -- the horizontal purple line indicates $\rest=1/\parallax$.
\label{fig:distest_summaries}
}
\end{center}
\end{figure}

Figure~\ref{fig:distest_summaries} summarizes the properties of the distance estimates for unimodal posteriors with mode and HDI estimates. We discuss this figure for most of the rest of this section.

Panel (a) shows the distribution of the distance estimate \rest, with the distribution of the length scale, \rlensph, in panel (d) beneath it for comparison.
Some idea of the fractional distance uncertainties is shown by the histogram in panel (b), where $(\rhi-\rlo)/2$ is taken as a single symmetrized measure of the uncertainty. This can be compared with the distribution of the fractional parallax uncertainties shown in Figure \ref{fig:parallax_distributions}b. The fractional distance uncertainties are generally smaller because the prior prevents the posterior distribution becoming arbitrarily wide.

Panel (c) indicates the skewnesses of the posteriors. The upper part of the confidence interval ($\rhi-\rest$) is always larger than the lower part ($\rest-\rlo$). In 2\%, 0.6\%, and 0.15\% of the cases it exceeds it by factors of more than 5, 10, and 20 respectively.  As mentioned earlier, if the asymmetry is so large that the positive tail is nearly flat, then the HDI may not have been estimated very accurately.

Panel (e) shows how the fractional distance uncertainty varies with distance. The complex shape is a consequence of the true distribution in the Galaxy, the variation of $\rlensph(\glon, \glat)$, and the properties of the posterior.

The relation between the fractional parallax uncertainty and distance uncertainty is shown in panel (f). For small (positive) values of both parameters they are highly correlated. For large positive values of \fpu, the fractional distance uncertainty can never get very large, due to the prior. For the prior itself, $(\rhi-\rlo)/(2\rmode)\simeq0.74$ (independent of \rlen), and this is the value that the upper envelope of the distribution asymptotes to as $\fpu \rightarrow \infty$.
For some intermediate values of \fpu\ the fractional distance uncertainties are quite large. These are due to posteriors with long positive tails like that shown in Figure~\ref{fig:posterior_prior_examples}(d): $\rhi \gg \rest$ which results in 
$(\rhi-\rlo)/2\rest$ being large. If we used the median or mean as a distance estimate instead, this measure of the fractional distance uncertainty would not extend to such large values (although those estimators would introduce other problems).

Panel (g) plots the ratio of the mode of the posterior (\rest) to the mode of the prior ($2\rlensph$) as a function of the fractional parallax uncertainty, \fpu, to give some sense of how dominant the prior is. 
The posterior is a (renormalized) product of the prior and likelihood, so
the location of the posterior mode depends not only on the modes of the prior and likelihood -- the latter is at 
$1/\parallax$ when considered as a function of distance -- but also on their widths, i.e.\ how peaked/flat the distributions are. This generates a rather complex behaviour in panel (g).
For small positive values of \fpu\ (below 0.1--0.2) we see a wide range of values of $\rest/2\rlensph$, as we would expect: there is little constraint from the prior. 
%
As \fpu\ increases, the prior plays more of a role and so the posterior mode deviates from it by less. 
We see a sharp lower boundary in the scatter plot, showing that \rest\ cannot drop below some finite multiple of the prior mode. This is due to the shape of the prior density function, which always drops to zero at zero distance: for a likelihood of finite width (i.e.\ $\fpu>0$), its product with the prior always produces a mode above zero.
The upper envelope, in contrast, is blurred because there is no hard upper limit on the location of the prior mode.

A comparison of the distances to the naive inverse parallax estimates is shown in panel (h)
by plotting their ratio (i.e.\ $\rest \parallax$), and zoomed in for the higher-precision parallaxes in panel (i). 
Once \fpu\ exceeds around 0.1--0.3, the inverse parallax deviates considerably from \rest.
For large fractional parallax uncertainty, \rest\ tends to be less than $1/\parallax$. This is because a noisy inverse parallax can extend to arbitrarily large distances, something that the prior constrains.
The inverse parallax is known to be a biased and very noisy estimator \citep[e.g.][]{2015PASP..127..994B,DR2-DPACP-38} which tends to overestimates the true distance.

\subsection{Spatial distribution}

\begin{figure}
\begin{center}
\includegraphics[width=\textwidth, angle=0]{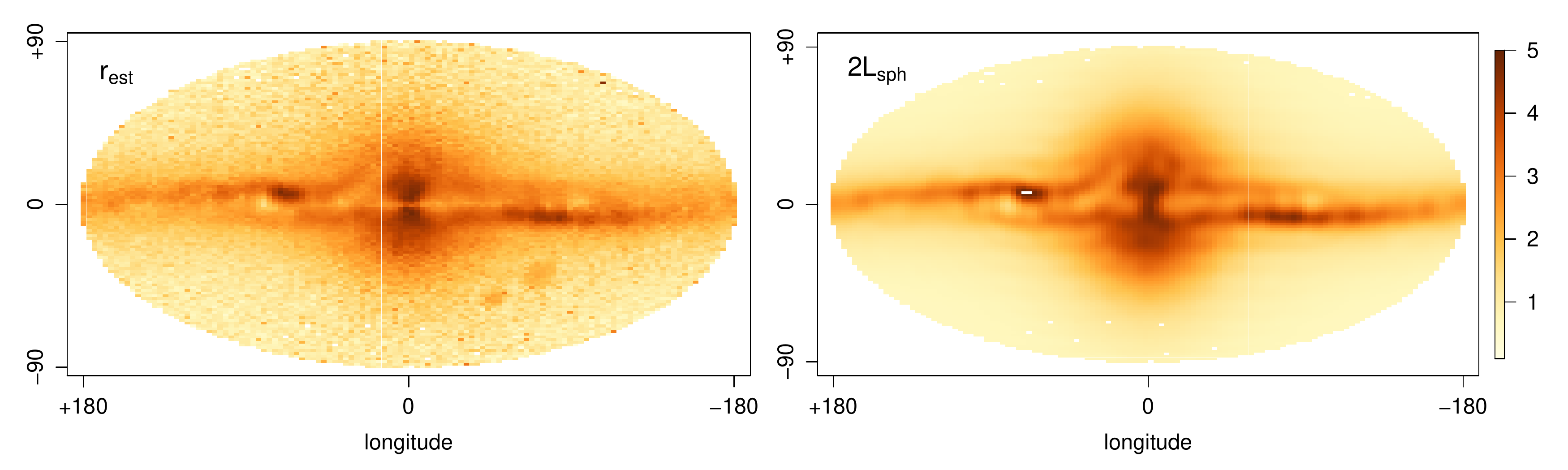}
\caption{Distribution of distances for sources with 
{\tt result\_flag=1} and {\tt modality\_flag=1} shown in Galactic coordinates on a Mollweide projection.
The colour scale (the same in both panels) shows the mean distance, in kpc on a linear scale, for sources over a small range of (\glon, \glat). White cells lie outside the range 0--5\,kpc.
Left: The estimated distances in the catalogue, \rest\ (i.e.\ the posterior mode). Right: The prior mode ($=2\rlensph$).
\label{fig:rest_2rlen_galproj}
}
\end{center}
\end{figure}

The left panel of Figure~\ref{fig:rest_2rlen_galproj} shows the distribution of estimated distances in Galactic coordinates, for those stars where the posterior is unimodal and this mode is the distance estimator. Compare this to the right panel, which shows the distribution of the modes of the corresponding priors. The overall features of the prior remain evident in the left panel, partly because such features correspond to real, previously known features in the Galaxy, but also because most stars have large fractional parallax uncertainties. Yet we clearly see features that are not present in the prior map, such as the Large and Small Magellanic Clouds.  Our mean distances to these satellite galaxies are of course underestimated, because 
the stars have very large fractional parallax uncertainties, so our estimates are prior-dominated.
The largest length scales in the prior model are only about 2.6\,kpc -- corresponding to prior modes of 5.2\,kpc -- whereas these galaxies are in fact about 50--60\,kpc away. The same problem applies to distant giants, because the prior will normally be dominated by the nearer dwarfs in the model.

\begin{figure}
\begin{center}
\includegraphics[width=\textwidth, angle=0]{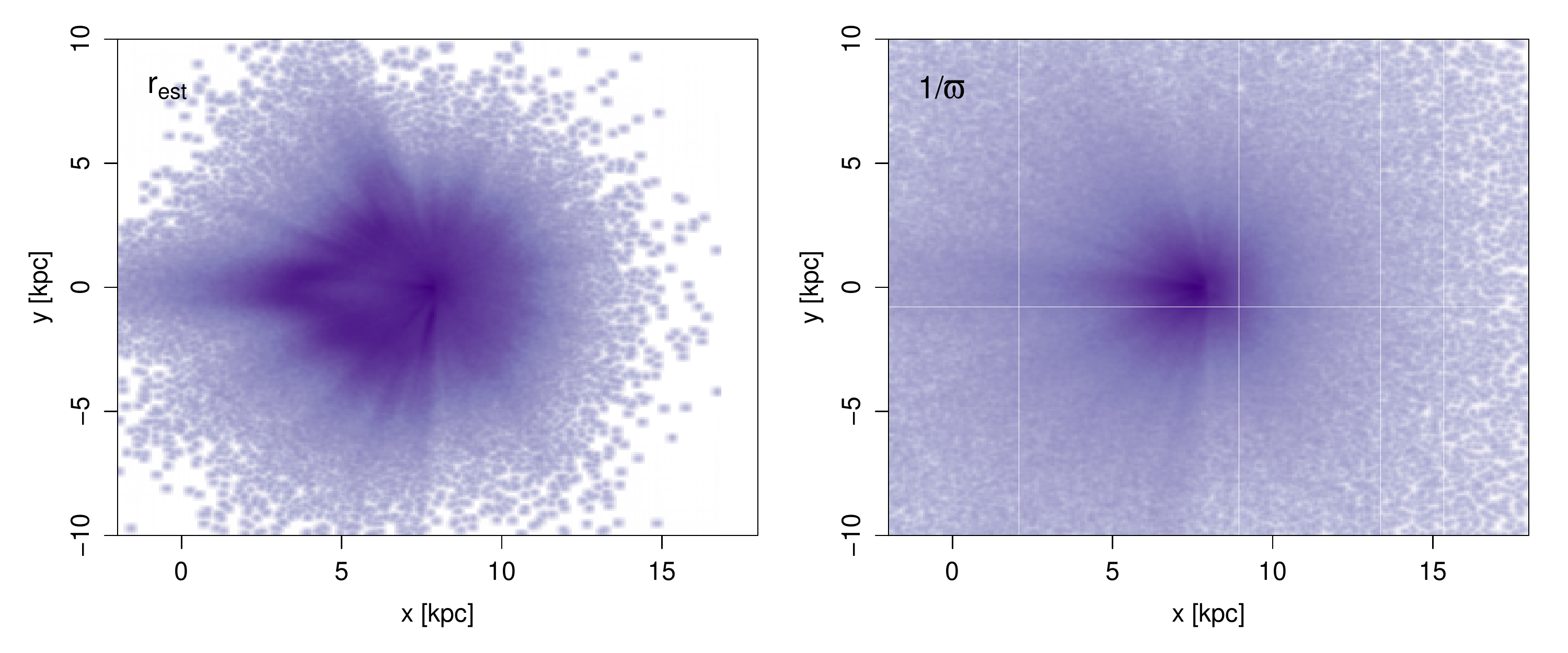}
\caption{Distribution of stars' positions projected onto the Galactic plane shown as a density plot. The Galactic Centre is at $(0,0)$\,kpc and \gaia\ (and the Sun) are at $(+8,0)$\,kpc. The left panel is constructed using our distance estimates (for {\tt result\_flag=1} and {\tt modality\_flag=1}). The right panel is constructed using the naive $1/\parallax$ distance estimator (with the additional requirement that $\parallax>0$).
\label{fig:galaxy_xy_rest}
}
\end{center}
\end{figure}

Figure~\ref{fig:galaxy_xy_rest} compares the projected distribution of stars in the Galactic plane from our distance estimates (left) with those obtained from a naive parallax inversion (right). This shows the tendency of inverse parallax to extend to implausibly large distances. The radial lines from the Sun in both panels are due to nearby dust clouds diminishing the number of stars observed along certain sight-lines.

\subsection{Validation using clusters}

\begin{figure}
\begin{center}
\includegraphics[width=\textwidth, angle=0]{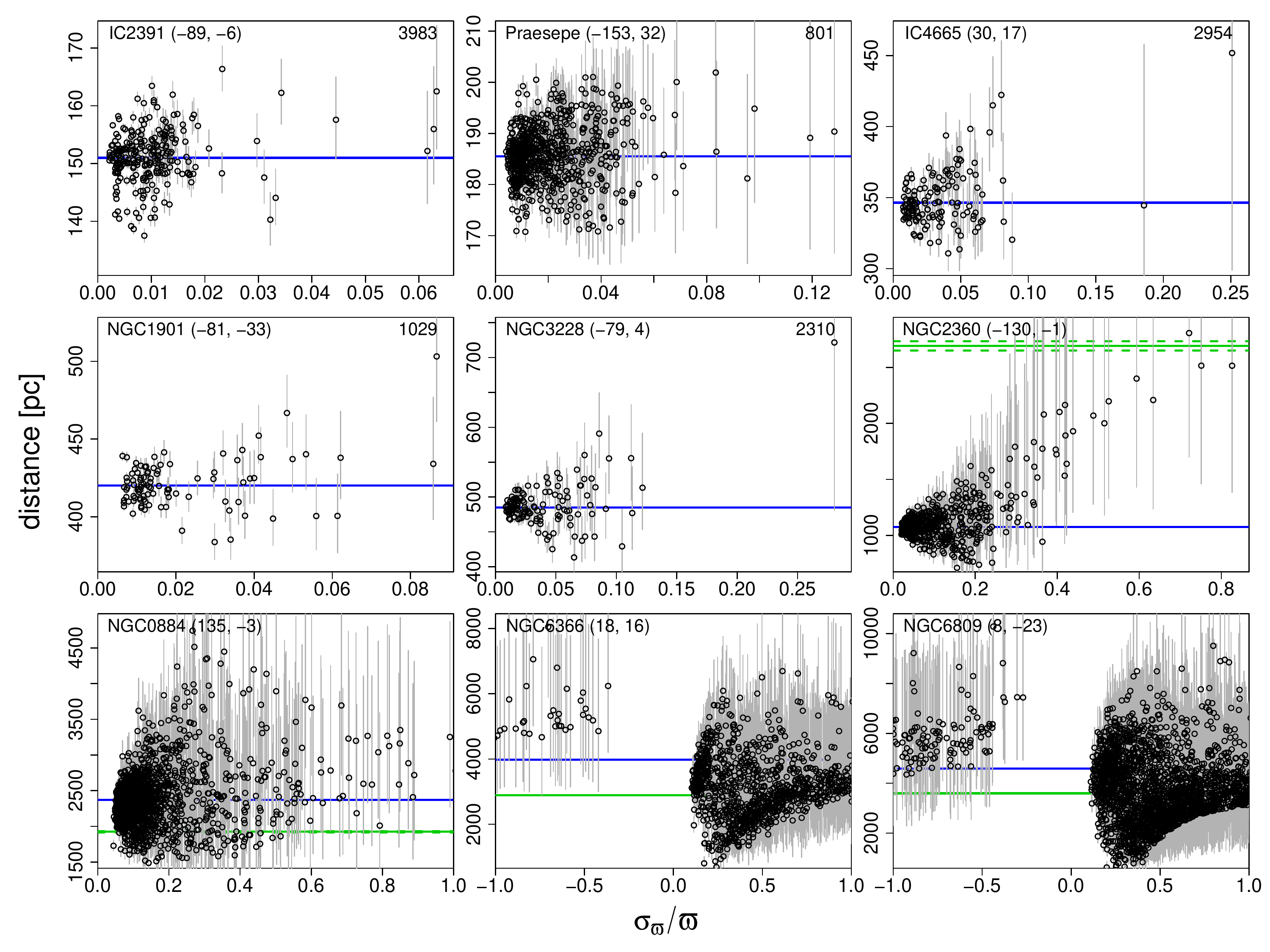}
\caption{Validation of distance estimates using clusters (one per panel, with the name and $(\glon,\glat)$ coordinates shown in the top left corner of each).
We plot the estimated distances, \rest, of the cluster members as open circles, as a function of \fpu. The error bars show the lower (\rlo) and upper (\rhi) bounds of the confidence interval for each star. 
The range of \fpu\ in each panel is from $\max(-1, \min(0,\fpu))$ to $\min(1,\max(\fpu))$.
The vertical range of each panel is set to span the full range of \rest\ for everything lying in the range of \fpu\ plotted.
The solid green horizontal line is $2\overline{\rlensph}$, the average of the modes of the priors for all cluster members. The dashed green lines around this (sometimes barely visible) show the standard deviation in this across the cluster members. In several cases these measures are outside the limit of the plot, in which case $2\overline{\rlensph}$ is written in the top right corner of the panel. The solid blue line is the inverse of the mean cluster parallax for all members (including any beyond the \fpu\ limits plotted). The clusters are ordered by increasing cluster distance.
\label{fig:cluster_validation}
}
\end{center}
\end{figure}
 
A simple validation of our distance estimates can be obtained using stellar clusters, on the assumption that the members of a physical (gravitationally bound) cluster span just a small range of distances.

Given a list of clusters with known sky positions, we select cluster members based on their proper motions. That is, we assume all members follow a common space motion (with some dispersion) that is different from the bulk of the field stars in the same region. We model the proper motion plane using Gaussian mixture models for both the cluster region and a field region. This allows us to assess the membership probability to the cluster on a star-by-star basis. For present purposes we focused on purity by selecting only the most probable members. This method gives very a similar selection of members to that found and reported in \cite{DR2-DPACP-31}.

Figure~\ref{fig:cluster_validation} shows some results of validating our distance estimates using clusters.
Each panel shows our distances to individual cluster members as a function of the fractional parallax uncertainty.
(In all cases the estimates turn out to have {\tt result\_flag=1} and {\tt modality\_flag=1}.)
These distances can be compared to the inverse of the mean parallax of the cluster members, shown as the blue line.
This mean is computed using all members regardless of \fpu\ (including negative parallaxes if present).
Although the parallax uncertainties are correlated on the angular scale of most stellar clusters ($\lesssim 1^\circ$), these correlations
need not be considered when forming a simple mean. They would need to be accommodated if we estimated the formal uncertainty in this mean. Yet due to the large number of cluster members, the uncertainty in the mean is instead limited by the systematic parallax error, which is generally below 0.1\,mas \citep{DR2-DPACP-51}. Thus for clusters within one kpc or so, the inverse mean parallax is a reasonable approximation of the cluster distance. We are not suggesting that this is the best way to estimate true cluster distances. We are just using it to give some baseline against which to compare our distance estimates (and uncertainties therein).

Figure~\ref{fig:cluster_validation} shows that for nearby clusters (first five panels), our distance estimates for the individual stars agree with the cluster mean. This is as expected, since for nearby clusters all the members have small \fpu, and for small \fpu\ the posterior mode is similar to inverse parallax.  In all of these cases the prior has negligible impact (the modes of the priors are much larger than the estimated distances).  These panels also show that the estimated uncertainties (the error bars plotted) are reasonable, considering that clusters have a finite depth.

As we move to more distant clusters, we see some informative behaviour. First, we see that some cluster members have large \fpu. As discussed earlier, the posteriors for stars with large \fpu\ are often dominated by the prior (e.g.\ Figure~\ref{fig:posterior_prior_examples}f). This is seen nicely in the panel for NGC2360, where the estimates move up towards the prior for large \fpu. The error bars also get larger.
The next cluster, NGC0884, doesn't show such a clear trend, and in fact at large \fpu\ the distance estimates are systematically larger than the prior. This is an example where even for poor parallaxes the posterior is not entirely dominated by the prior, because the likelihood, even though broad, suggests quite a different distance than the prior does. This was also seen in Figure~\ref{fig:posterior_prior_examples}g. The final two clusters are dominated by stars with large \fpu. Here the posteriors tend to be closer to the priors. Both clusters are so distant that many of their members have negative parallaxes. Yet our inferred distances to these are broadly consistent with the mean cluster parallax. Large uncertainties notwithstanding, this demonstrates that useful information can be extracted from negative parallaxes. Of course for these more distant clusters, the mean cluster parallax has a significant systematic error (order 0.1\,mas), as do the individual parallaxes used for our distance estimates.

\section{Distance catalogue content and characteristics}\label{sec:catalogue}

\begin{table}
\begin{center}
\caption{The format of the distance catalogue, showing data on five sources chosen at random. 
The source\_id is a long integer, and is the same as that in \gdr{2}.
\rest\ is the point distance estimate. \rlo\ and \rhi\ are the lower and upper bounds of the (approximately) 68\% confidence interval.
\rlensph\ is the length scale of the prior used.
Distances are shown rounded to 0.001\,pc here, but are given at full numerical precision in the catalogue.
{\tt result\_flag} can take values: 0 (no solution, so \rest, \rlo, and \rhi\ are all {\tt NaN}); 1 (\rest\ is the mode, and \rlo\ and \rhi\ define the highest density interval, HDI); 2 (\rest\ is the median, and \rlo\ and \rhi\ define the equal-tailed interval, ETI). {\tt modality\_flag} gives the number of modes of the posterior: it is either 1 or 2. If 
 {\tt result\_flag=1} for bimodal posteriors, the mode reported is the
 highest mode, and the highest density interval is defined around
 this.
\label{tab:catalogue_extract}
}
\begin{tabular}{rrrrrrr}
\hline
source\_id & \rest & \rlo & \rhi & \rlensph & result & modality \\
                 & pc & pc & pc & pc & flag & flag \\
\hline
6056824904455202560 & 2471.511 & 1144.487 & 4870.877 & 1324.036 & 1 & 1 \\
3125855551398349952 & 2914.102 & 1456.329 & 5343.421 & 1307.715 & 1 & 1 \\
5545891471750684160 & 3672.499 & 2358.817 & 5960.471 & 1340.001 & 1 & 1 \\
1824805611208672256 & 2919.521 & 1338.509 & 5546.599 & 1407.083 & 1 & 1 \\
4663657669625864448 & 1026.809 & 583.985 & 1896.781 & 504.083 & 1 & 1\\
\hline
\end{tabular}
\end{center}
\end{table}

\begin{table}
\begin{center}
\caption{Minimum and maximum values of the seven columns in the distance catalogue, as well as the 1st, 50th, and 99th quantiles of the four distance columns.
For all sources $\rhi > \rest > \rlo$. \rlo\ can be exactly zero. 
The smallest values of \rest\ are almost certainly the result of spurious parallaxes.
\label{tab:catminmax}
}
\begin{tabular}{rrrrrrrr}
\hline
&     {\tt source\_id}                                 &      {\tt r\_est}  &      {\tt r\_lo}   &      {\tt r\_hi}                &     {\tt r\_len}  &   {\tt result\_flag} & {\tt modality\_flag} \\
minimum & 4295806720                        &   0.540           &    0.000            &    0.540                       & 335.227         & 0 & 1 \\          
1st percentile   & &  318.91      &   238.373   &   397.475       & 384.145   & & \\
50th percentile & & 2841.107 &   1572.914 &   5193.845    & 1473.436   & & \\
99th percentile & & 7317.404   &   5074.212 &   11\,169.098 & 2388.871   & & \\
maximum & 6917528997577384320    & 32\,978.685  & 27\,577.405   & 155\,560\,896.912    & 2597.694       & 2 & 2 \\ 
any {\tt NaN}?  & no & yes & yes & yes & no & no & no \\
\hline
\end{tabular}
\end{center}
\end{table}

\begin{table}
\begin{center}
\caption{The numbers of sources with different combinations of the flags.
The total number of sources is 1\,331\,909\,727.
HDI = highest density interval; ETI = equal-tailed interval.
\label{tab:summary_statistics}
}
\begin{tabular}{rrrrr}
\hline
& & \multicolumn{2}{r}{modality} & \\
& & \multicolumn{2}{r}{flag} & \\
& & {\tt 1} & {\tt 2} & Sums\\
\cline{3-4}
\multirow{3}{*}{result flag} 
& (fail)             {\tt 0} &               3278        &                 0 & 3278 \\
& (mode+HDI) {\tt 1} & 1\,330\,715\,981 &   295\,451 & 1\,331\,011\,432\\
& (median+ETI) {\tt 2} &            13\,477     & 881\,540 & 895\,017\\
\cline{3-4}
\multicolumn{2}{r}{Sums} & 1\,330\,732\,736 & 1\,176\,991 & \\
\hline
\end{tabular}
\end{center}
\end{table}

The distance catalogue comprises 1\,331\,909\,727 entries with seven columns. An extract for five sources selected at random, together with the definition of the columns, is given in Table~\ref{tab:catalogue_extract}.
The minimum and maximum values of the columns are given in Table~\ref{tab:catminmax}.
Table~\ref{tab:summary_statistics} summarizes the six different types of solutions possible, corresponding to the six combinations of the two flags.
The vast majority of posteriors are unimodal and could be successfully summarized with the mode and HDI
({\tt result\_flag=1}, {\tt modality\_flag=1}).
A very small fraction are instead summarized with the median and ETI
({\tt result\_flag=2}, {\tt modality\_flag=1}). 
Some bimodal posteriors are summarized with the mode and HDI
({\tt result\_flag=1}, {\tt modality\_flag=2}),
although for most the HDI is not uniquely defined, so the median and ETI are reported instead
({\tt result\_flag=2}, {\tt modality\_flag=2}).

As noted before, in a very small fraction of cases the posterior is so skew that the computed HDI actually spans a probability that is much larger than 0.6827 (but not as high as 0.9, which is what forces the solution to {\tt result\_flag=2}). 
Such posteriors can be identified from the fact that then $\rhi - \rest \gg \rest -\rlo$.

When doing statistical analyses on an ensemble of the catalogue results one should not mix the two types of solution (values of {\tt result\_flag}), as the estimates are conceptually different.
 If one is interested in using the bimodal solutions, we strongly suggest one plots the posteriors and decide whether the summaries provided are appropriate. Note that adopting a different length scale could remove the bimodality.

No post-processing has been performed to modify distances or to remove entries which could potentially be identified as implausible based on additional information (such as a source classification, photometry, or proper motions). 
\gdr{2} is known to contain a number of sources with spuriously large parallaxes, some of which are (spuriously) precise (e.g.\ 174 sources with $\parallax>500$\,mas, all of which have $\fpu<0.01$).
This produces, for example, 21 sources in our distance catalogue with $\rest<1$\,pc (they all have $\gmag>19$). 
While they could in principle be nearby brown dwarfs (and many have high proper motions), they are probably cross matching errors in the data processing. It is very unlikely that \gaia\ has discovered so many new nearby sources.
More conservative filtering described in Appendix C of \cite{DR2-DPACP-51} preferentially removes such sources, suggesting most are indeed spurious.
\cite{2018arXiv180409375A} (section 4.1 and Figure 10) show that spuriously large parallaxes are generally concentrated in regions of high stellar density.

The probability density of our prior reaches zero only asymptotically at large distances, so in principle extragalactic objects can be assigned arbitrarily large distances corresponding to their very small parallaxes. In practice, however, the half million or so extragalactic objects in \gdr{2} have statistically insignificant parallaxes, so the prior will dominate their distance estimates, meaning that these will be woefully too small.  
The same applies for stars in the local group.

Just as the \gaia\ astrometric data processing treats all sources as single stars moving on linear paths relative to the Sun, so our distance estimates ignore any such binarity.

The individual distance estimates are coupled on angular scales on the sky through the use of a smooth model for the prior length scales, $\rlensph(\glon,\glat)$. This is important when the prior dominates the posterior, because the combination of multiple distance estimates (or even the entire posterior) for such stars will compound the prior.
As described in \cite{DR2-DPACP-51}, the parallax uncertainties show between-source correlations on various angular scales on the sky, particularly below 1 degree. Our distance uncertainties will be similarly correlated.
This must be considered when combining data, for example to estimate the distance to a cluster based on several of its members. We do not recommend using distance estimates in such combinations, however. It is better to make a model of the cluster and infer its distance and size simultaneously. A tutorial by one of us (CBJ) describing a simple approach to this can be found in \cite{DR2-DPACP-38}.

While it would have been convenient for the user if these distances were part of the main \gdr{2} catalogue, it was decided by the DPAC Executive not to include any distances in the official release. Our distance catalogue is nonetheless available on the Gaia archive \url{http://gea.esac.esa.int/archive/} as an external catalogue table called
{\tt  external.gaiadr2\_geometric\_distance}.
It is therefore easy to make queries which join the two tables on the source identifier.  The following ADQL query gives an example of how to do this by selecting nearby hot stars in a single healpix at level 4.
\begin{verbatim}
SELECT source_id, ra, dec, phot_g_mean_mag, r_est, r_lo, r_hi, teff_val
FROM external.gaiadr2_geometric_distance
JOIN gaiadr2.gaia_source USING (source_id)
WHERE r_est < 1000 AND teff_val > 7000
AND source_id < 2251799813685248
\end{verbatim}

\section{Summary and conclusions}\label{sec:summary}

We have produced a catalogue of distance estimates for 1.33 billion sources in the second \gaia\ data release.  By using a inference approach we can infer meaningful distances even when the parallaxes are negative and/or the parallax precision is low. These are in fact the most common cases in the \gaia\ catalogue.  The estimation of the distance to each source involves a prior with a single length scale parameter.  This parameter, obtained by fitting to a three-dimensional model of the Galaxy as seen by \gaia, varies smoothly and realistically as function of Galactic longitude and latitude (only).  In the limit of precise parallaxes, say better than about 10\% (77 million sources),
the prior has little impact on the resulting distance estimates.  As the parallax precision degrades, the information from the measurement becomes poorer, and our background information as represented by the prior starts to take over.  The advantage of the inference approach is that as it ensures a smooth and graceful transition between the data-dominated and prior-dominated regimes, thus assuring self-consistency over the entire catalogue. This is important, because selecting sources with small fractional parallax uncertainties
 -- perhaps in the hope of using inverse parallax as a distance estimator -- will create a sample biased toward bright nearby sources.

The other advantage of the probabilistic approach adopted here is that we also infer a sensible confidence interval on every distance estimate. These must not be ignored, as there is a genuine and unavoidable uncertainty in the distance estimates. We specify this confidence interval via its lower and upper bounds. These are asymmetric with respect to the distance estimate on account of the nonlinear transformation from parallax to distance.

We remind the reader that it was our explicit intention to provide purely geometric distances, free from assumptions of stellar physics or interstellar extinction toward individual sources. We believe such distances and their uncertainties are useful for various cases such as those described in the introduction.
We nonetheless realise that more precise -- if not necessarily more accurate -- distances can be estimated for a targeted subset of stars for which we use more information. Finally, we should emphasise that there are many tasks where explicit distances should not be used at all. If the distance is only an intermediate step in an analysis, for example when estimating luminosities, or
transverse velocities from the proper motions, you will almost always want to do the inference from first principles, working directly with the parallaxes.

\acknowledgments

This work was funded in part by the DLR (German space agency) via grant 50 QG 1403.
It has made use of data from the European Space Agency (ESA) mission \gaia\ (\url{http://www.cosmos.esa.int/gaia}), processed by the \gaia\ Data Processing and Analysis Consortium (DPAC, \url{http://www.cosmos.esa.int/web/gaia/dpac/consortium}). Funding for the DPAC has been provided by national institutions, in particular the institutions participating in the \gaia\ Multilateral Agreement. We thank the IT departments at MPIA and ARI for computing support.

\bibliographystyle{aasjournal}
\bibliography{distances,gaia}

\end{document}